\documentclass[11pt,a4paper]{article}
\pdfoutput=1
\usepackage{a4wide}
\usepackage{amsfonts}
\usepackage{amssymb}
\usepackage{amsmath}
\usepackage{graphicx}
\usepackage{multirow}
\usepackage{float}
\usepackage{booktabs}
\usepackage{array}
\usepackage{color}
\usepackage{ulem}
\usepackage[small,bf]{caption}
\usepackage{cite}
\usepackage[usenames,dvipsnames]{xcolor}
\DeclareMathOperator{\erf}{erf}
\usepackage{verbatim}
\usepackage{textcomp}
\usepackage{caption}
\usepackage{subcaption}

\allowdisplaybreaks

\numberwithin{equation}{section}

\definecolor{DeepPink}{rgb}{1.0, 0.08, 0.58}

\definecolor{Blue}{rgb}{0.08, 0.08, 1.00}

\definecolor{Green}{rgb}{0.0, 0.8, 0.24}

\definecolor{candyapplered}{rgb}{1.0, 0.03, 0.0}

\newcommand{\DM}{{\text{DM}}}
\newcommand{\SNR}{\rho^2_{\text{KAGRA}}}

\DeclareMathOperator{\diff}{\text{d}}

\begin{document}
\begin{titlepage}


\begin{center}
{
\bf\LARGE Discovery prospects for heavy dark matter\\[0.2em] in KAGRA
}
\\[8mm]
Chun-Hao Lee$^{\, a,}$ \footnote{E-mail: \texttt{chlee@gapp.nthu.edu.tw}},
Reinard Primulando$^{\, b,}$ \footnote{E-mail: \texttt{rprimulando@unpar.ac.id}} and 
Martin Spinrath$^{\, a,\, c}$ \footnote{E-mail: \texttt{spinrath@phys.nthu.edu.tw}}
\\[1mm]
\end{center}
\vspace*{0.50cm}
\centerline{$^{a}$ \it Department of Physics, National Tsing Hua University, Hsinchu 30013, Taiwan}
\vspace*{0.2cm}
\centerline{$^{b}$ \it Center for Theoretical Physics, Department of Physics}
\centerline{\it Parahyangan Catholic University, Jalan Ciumbuleuit 94, Bandung 40141, Indonesia}
\vspace*{0.2cm}
\centerline{$^{c}$ \it Center for Theory and Computation,
	National Tsing Hua University, Hsinchu 30013, Taiwan}
\vspace*{1.20cm}

\begin{abstract}
\noindent
We discuss the discovery prospects for kg-scale dark matter with a Yukawa-like long-range 
interaction in Kamioka Gravitational Wave Detector (KAGRA). We consider the interaction 
range to be in the order of kilometers, thus the dark matter may interact with multiple KAGRA mirrors simultaneously. The induced signal strain is in many cases dominated by frequencies below about 130 Hz. Due to the geometry of the detector there is some strong directional sensitivity. It turns out that KAGRA would be able to constrain such kind of dark matter within a few years of operation. With our set of assumptions we expect that KAGRA could detect a few events per year for 10 kg dark matter, with a Yukawa interaction with a range of 10 km and an effective coupling constant around $10^{6}$.
\end{abstract}

\end{titlepage}
\setcounter{footnote}{0}

\section{INTRODUCTION}
\label{sec:Introduction}

Despite dominating the matter density of the universe, little is known about the nature of dark 
matter (DM). The lower bound of the dark matter mass is about $10^{-22}$ eV while the upper 
bound is around five solar masses~\cite{ParticleDataGroup:2020ssz}.
Given that there is an 80 orders of magnitude difference between the lower and upper bound 
of this mass range, many different possible experiments need to be considered in unraveling 
the properties of dark matter. Many current direct detection experiments have explored dark 
matter in the WIMP mass window around 100 GeV and axion dark matter with masses well 
below 1 eV. These mass regions are motivated by the solutions to other problems. Primordial 
black holes with masses heavier than asteroid masses have also recently gained attention as 
heavy dark matter candidates. But all direct searches have not found any conclusive evidence 
so far. Large regions of model parameter space have been constrained 
(see Refs.~\cite{ParticleDataGroup:2020ssz, Schumann:2019eaa, Chadha-Day:2021szb, Villanueva-Domingo:2021spv}), 
but there are still an enormous mass and parameter range left which have been weakly 
explored, if at all. Hence, new ways to look for dark matter focusing on other mass regions 
and detection techniques have recently gained more attention and we will also go in this 
direction here.

To be precise, we want to study here the possibility of using a gravitational wave detector as 
a DM detector. There have been proposals going in a similar direction using either gravitational 
wave detectors or generally speaking optomechanical experiments~\cite{Kawasaki:2018xak, Hall:2016usm, Seto:2004zu, Adams:2004pk,Dimopoulos:2008hx, Dimopoulos:2008sv, Graham:2012sy, Riedel:2012ur,Stadnik:2014tta, Arvanitaki:2015iga, Stadnik:2015xbn, Graham:2016plp,Branca:2016rez, Riedel:2016acj, Graham:2017pmn, Canuel:2017rrp, Jung:2017flg,Pierce:2018xmy, Morisaki:2018htj, Coleman:2018ozp, Grabowska:2018lnd,Canuel:2019abg, Zhan:2019quq, Cheng:2019vwy, Badurina:2019hst, Bertoldi:2019tck,  Grote:2019uvn, PhysRevLett.122.223601, Monteiro:2020wcb, Jaeckel:2020mqa,Lee:2020dcd,Chen:2021apc,Baum:2022duc,Lu:2022zpn, Vermeulen:2020djm, Miller:2020vsl}.
However, the mass range we consider is very different from many of these proposals and 
hence we will use a different formalism.

We consider here one of the possible DM candidates in the kg- to ton-scale mass range. The 
DM in this mass range interacts significantly with the proposed detector while we can still have 
a detectable rate. This class of DM can be motivated by several models such as primordial 
extremal black holes~\cite{Bai:2019zcd}, dark quark nuggets~\cite{Bai:2018dxf}, Fermi 
balls~\cite{Ogure:2002hv} and electroweak symmetric DM balls~\cite{Ponton:2019hux}. 
Currently, there is no dedicated direct detection search for this class of the dark matter to our 
knowledge. There have been some proposed searches sensitive to such DM using an array 
of force sensors~\cite{Carney:2019pza,Blanco:2021yiy} or precision displacement
sensors~\cite{Kawasaki:2018xak}. It is also possible that dark matter can have an additional
Yukawa-like interaction with the normal matter besides the gravitational interaction.
What will be of particular importance here is that for a longer range interaction, the dark matter 
might interact simultaneously with two or more different objects located in some distance. Therefore 
correlations between the different objects need to be taken into account. This can lead to some 
interesting phenomena, such as signal enhancement or suppression and directional sensitivity. 
This long-range interaction also implies that dark matter can have very large cross sections for 
interactions with ordinary matter compensating the very low number density.

Our work here closely follows the model and general proposal in~\cite{Hall:2016usm} which 
discussed the prospects for looking for heavy DM in LIGO. But we will go significantly beyond 
that adding details on the modeling of the signal-to- noise (SNR) ratio, including a more realistic 
noise model, and discussing more explicitly various parameter depend- encies. The work here 
relies heavily also on the insights from~\cite{Lee:2020dcd} by some of us where we modeled 
DM scattering in KAGRA. The difference is that in this previous work we considered light DM 
with a very short-range interaction. In our work here instead, we will consider a kg- to ton-scale 
DM with $\mathcal{O}$(km)-range interaction with the normal matter.
This suggests that we look for a small, but correlated force on objects which are also about 
$\mathcal{O}$(km) away from each other. Ideal detectors for that are terrestrial gravitational 
wave detectors which can detect minuscule relative dis- placements of their mirrors arranged in 
an interferometer. We consider here as an explicit example the currently running KAGRA 
observatory. KAGRA is the most recent, large scale gravitational wave detector with the most 
modern technology making it an ideal benchmark for the near future as well. We use the 
realistic KAGRA suspension setup and noise estimated by the KAGRA collaboration in this 
work to provide a reasonable benchmark.

Our paper is organized as follows: In Sec.~\ref{sec:Framework}, we explain how to calculate 
the observable for a DM signal in KAGRA and similar experiments, starting from the potential 
which consists of a Yukawa and a Newtonian term. With this information and having a noise 
model at hand we discuss how to calculate the SNR. In Sec.~\ref{sec:Directions}, we then 
show the dependence of the SNR on the DM phase space and other dark sector parameters. 
In Sec.~\ref{sec:Rates}, we estimate the event rate per year for various DM masses, the 
Yukawa range, and the coupling strengths. We summarize and conclude this work in 
Sec.~\ref{sec:Summary}.

\section{MODELING HEAVY DM SCATTERING IN KAGRA}
\label{sec:Framework}

In this paper we assume that DM has a gravitational and a long-range Yukawa interaction 
with the mirrors in KAGRA. The potential for this interaction with one of the test masses is, 
cf.~\cite{Hall:2016usm},
\begin{align}
\label{eq:DMpotential}
V_\DM = - m_T m_\DM \frac{G_\text{N}}{|\vec{x}_T - \vec{x}_\DM|} \left( 1 + s \, \delta_{\text{SM}} \delta_\DM \exp(-|\vec{x}_T - \vec{x}_\DM|/\lambda ) \right) \;,
\end{align}
where $\vec{x}_{T}$, $\vec{x}_{\DM}$ and $m_{T}$, $m_{\DM}$
are position and mass of the target and the passing DM object respectively. 
The constants $\delta_{\text{SM}}$ and $\delta_\DM$ are effective coupling constants.
$G_\text{N}$ is Newton's constant and $\lambda$ parametrizes the range
of the Yukawa interaction. The factor $s = +1$ for a scalar
and tensor exchange and $-1$ for vector exchange. Throughout the paper
the Yukawa part will be dominant and we set $s = +1$. The differences
to $ s = -1$ are marginal in all numerical results in this paper. 

We assume
\begin{align}
 r_\DM , r_T  \ll L , \lambda \ll d_\DM \;,
\end{align}
where $r_\DM$ and $r_T$ measure the size of the DM object and the target, respectively,
and $L$ is the length of the interferometer arms. While $d_\DM$ is defined as 
$d_\DM^{-3} \equiv n_\DM$ in which $n_\DM$ is the local DM number density. This implies 
that we can approximate the DM object and the targets as
pointlike objects.

In this work we will focus on DM masses between 0.1 and 1000 kg, ranges of Yukawa forces 
between 0.1 and 10 km, and an effective SM-DM coupling $|\delta_\text{SM} \, \delta_\DM|$
between $10^{3}$ and $10^{7}$. One could be concerned at this point with constraints from 
fifth force searches. Indeed, if this would be an SM-SM force, that would be ruled out. There 
the Yukawa coupling $|\delta_\text{SM}|^2$ is constrained to be less than about 
5$\times 10^{-4}$~\cite{Adelberger:2009zz} for the considered range in $\lambda$.

But as it was also discussed, for instance, in \cite{Hall:2016usm} and \cite{Baum:2022duc}, 
this bound does not directly apply here but it can be combined with a limit on the cross section 
of DM self-interactions,
$\sigma_T/m_\DM \lesssim 1$~cm$^2$/g at $v_\DM \sim 1000$~km/s. In our notation
\begin{align}
 \frac{\sigma_T}{m_\DM} \sim
    \frac{16 \, \pi \, G_N^2 \, m_\DM \, |\delta_\DM|^4}{ v_\DM^4} \log\left( \frac{\lambda}{r_\DM} \right)
\end{align}
and
\begin{align}
  |\delta_\DM| &\lesssim 
    1.7 \times 10^{10} \frac{\text{kg}^{1/4}}{m_\DM^{1/4}} \frac{v_\DM}{1000 \text{ km/s}} \frac{(1 \text{ cm}^2/\text{g})^{1/4}}{(\sigma_T/m_\DM)^{1/4}} \;,
\end{align}
where we have used $\log\left( \lambda/r_\DM \right) = 5$.
Hence, after combining the fifth force and DM self-interactions constraint
$|\delta_\DM \delta_\text{SM}|\lesssim 10^8$ and our considered range
 $|\delta_{\text{SM}} \delta_\DM|$ between $10^3$
and $10^7$ is below current bounds.

Since the realistic forces are tiny, we neglect the back-reacting force on the DM object and 
hence its velocity $\vec{v}_{\DM}$ is constant,
\begin{equation}
 \vec{x}_\DM(t) = \vec{v}_{\DM} \, t + \vec{x}_{0,\DM} \;,
\end{equation}
where $\vec{x}_{0,\DM} = \vec{x}_\DM$ at $t = 0$ is the initial position
of the DM particle. Of course,
what time we choose to be $t=0$ is arbitrary and we will choose it such that
$|\vec{x}_{0,\DM}|$ is minimal in the frame with the corner of the interferometer
at the origin. If there would be only one target mass at the origin $|\vec{x}_{0,\DM}|$ 
would be very close to the impact parameter of the scattering process.
Please note that in the following we assume the DM velocity to be defined
in the lab frame and write $\vec{v}_{\DM}$ = $\vec{v}^{\, \text{lab}}_{\DM}$. This
will become important when we discuss the assumed DM velocity distribution which is usually
defined in the galactic rest frame.

KAGRA is an L-shaped interferometer with four mirror suspension systems located at both 
ends of each interferometer arm, cf.~Fig.~1 of \cite{KAGRA:2020tym}.
We choose one of the arms to be along the $x$-direction of our lab frame and the other one 
to be along the $y$-direction.
Their respective positions are labeled as EX, IX, EY, and IY [external (E)/internal (I) and 
$x$-/$y$-direction (X/Y)]. 
Each suspension system can be described as a system of three coupled, damped, 
harmonic oscillators, which are called intermediate mass (IM), blade springs (BS), and test 
mass or target mirror (TM) ~\cite{KAGRAnoise, Lee:2020dcd}.
As shown in~\cite{Chen:2016} and based on the experimental setup information
given in~\cite{KAGRAnoise} in each system the TM is at the bottom, the BS are on top
30~cm above the TM and the IM is 2~cm below the BS.
From now on, we will use indices $\textit{i} = 1$, $2$, $3$,
to refer to IM, BS, and TM respectively. 

Let us first consider the inner suspension system in the arm pointing
in $x$-direction (IX). Hence, we consider here the horizontal
external force $\vec{F}_{\text{ext},h}$ to point in $x$-direction as well. The equations of 
motion for this suspension system are given as
\begin{align}
	\left( M \frac{\diff^{2}} {\diff t^{2}} + K_{h} \right) \frac{\vec{x}_{h}(t)}{L} = \frac{\vec{F}_{\text{ext},h}(t)}{L} ,
	\label{eq:htheom1}
\end{align} 
where 
\begin{align}
	\label{eq:mkh}
	M =\begin{pmatrix}
		m_{1} & 0
		& 0  \\
		0
		& 
		m_{2} 
		&  0 \\
		0  & 0 &  m_{3}  
	\end{pmatrix}  \;&, \quad
	K_{h} = \begin{pmatrix}
		K_{1h}+K_{2h} & -K_{2h}
		& 0  \\
		-K_{2h}
		& 
		K_{2h}+K_{3h} 
		&  -K_{3h} \\
		0  & -K_{3h} &  K_{3h}  
	\end{pmatrix} \;, \\
	\vec{x}_{h}(t) =\begin{pmatrix}
		\breve{x}_{1h}(t) \\
		\breve{x}_{2h}(t) \\
		\breve{x}_{3h}(t)
	\end{pmatrix} &\text{ and }
	\vec{F}_{\text{ext},h}(t) = \begin{pmatrix}
		\breve{F}_{\text{ext},1h}(t) \\
		\breve{F}_{\text{ext},2h}(t) \\
		\breve{F}_{\text{ext},3h}(t)
	\end{pmatrix} \;.
	\label{eq:xF}
\end{align}
Note that the three components stand here for the components of the
suspension system, not the three directions of the lab frame.
The displacements $\breve{x}_{ih}$ are
 defined as the deviations in the lab frame $x$-direction
 from the respective equilibrium positions depending
 on the position of the mounting components. That means
 the total displacement of the IM in the lab frame
in $x$-direction is $\breve{x}_{1h}$, the total displacement of the BS is
$\breve{x}_{1h} + \breve{x}_{2h}$ and for the TM it is 
$\breve{x}_{1h} + \breve{x}_{2h} + \breve{x}_{3h}$.

The spring constants $K_{ih} \equiv k_{ih}(1+\text{i} \, 
\phi_{ih})$ are complex, for their values please see Ref.~\cite{KAGRAnoise}.
The masses of the suspension system components are $m_1 = 20.5$~kg, $m_2 = 0.22$~kg 
and $m_3 = 22.8$~kg.

One also has to consider the forces in vertical direction (the $z$-direction of our lab frame).
The form of the equation of motion for vertical direction would be 
the same but the values for the spring constants are different.
For more details, see Refs.~\cite{KAGRAnoise, Lee:2020dcd}.

In both cases, we
consider as external force the long-range force induced by a passing DM object, i.e.,
$\breve{F}_{\text{ext},ih}(t) = - \partial V_{\DM,i}/\partial \breve{x}_{ih}$ for the horizontal $x$-direction
and $\breve{F}_{\text{ext},iv}(t) = - \partial V_{\DM,i}/\partial \breve{x}_{iv}$ for the vertical 
$z$-direction,
where we have to use the respective positions of the suspension components in the potential.

We can take the Fourier transform of the equations of motion above and find
\begin{equation}
	\tilde{x}_{ih}(f) = \sum_{j=1}^{3}  \left[ \left( - (4 \pi^2) f^2 M  + K_{h}(f) \right)^{-1} \right]_{i j} \frac{\tilde{F}_{j,h}(f)}{L}
	\label{eq:tildex} \;,
\end{equation}
where we label Fourier transforms of quantities with a tilde and we have also absorbed
a factor of $L$ to make the displacement in frequency-domain dimensionless.

Since the coordinates $x_{ih}$ in Eq.~\eqref{eq:htheom1} are defined
relative to its predecessor, the Fourier transform of the TM $x$-position is given by
\begin{equation}
	\tilde{x}_{h}(f) = \tilde{x}_{3h}(f) + \tilde{x}_{2h}(f) 
	+ \tilde{x}_{1h}(f) \;
\end{equation} 
The other directions are completely analogous. The so-called horizontal direction 
points in the $y$-direction of the chosen frame for the other interferometer arm.

It is estimated that the vertical modes cross-couple to the horizontal modes due the 
curvature of the earth and mechanical imperfections~\cite{ALIGO:2001}. Hence, 
the relevant quantity from a single suspension system is 
\begin{equation}
	|\tilde{x}_{\DM}(f)|^{2} = (\text{VHC}^2 \times |\tilde{x}_{v}(f)|^{2}) + |\tilde{x}_{h}(f)|^{2}
	\label{eq:tot_displ}
\end{equation}
where VHC~$ = 1/200$ is the vertical-horizontal coupling.

\begin{figure}
	\centering
	\includegraphics[width=0.9\linewidth]{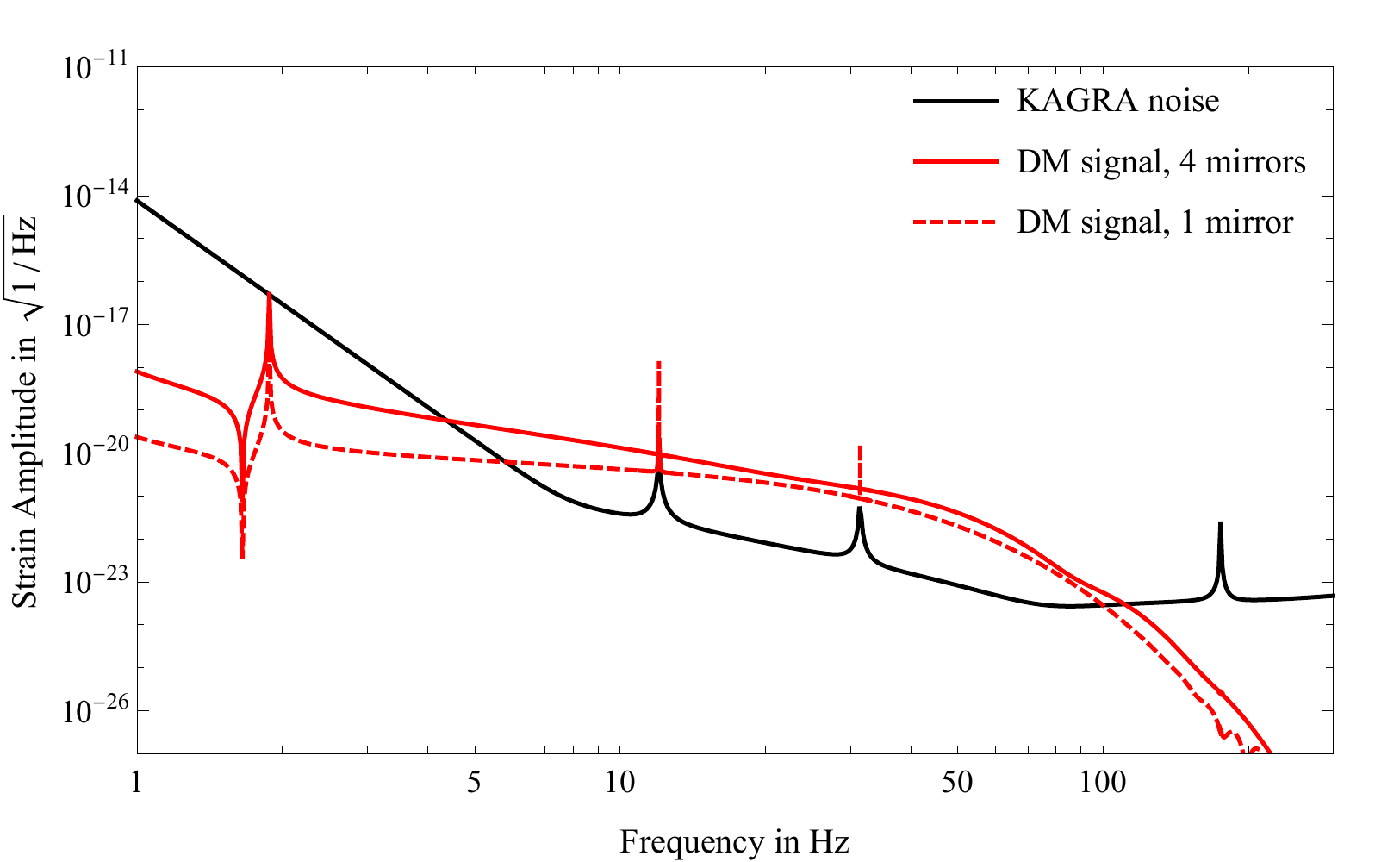}
	\caption{The comparison between the strain amplitudes of the DM signal of
		single suspension system (at IX location) and that of the four-suspension
		setup in KAGRA. The black line is the expected KAGRA background noise (see Ref.~\cite{KAGRAnoise}).
		For more explanations, see main text.
	}
	\label{fig:4mirror}
\end{figure}

The strain amplitude for just one dynamical suspension system 
is then given by
\begin{equation}
h_{\DM}(f) = \sqrt{4 f} |\tilde{x}_{\DM}(f)| \;.
\label{eq:dmstrain}
\end{equation}
In Fig.~\ref{fig:4mirror}, we plot it as a red dashed line,
where we chose for illustration, for the dark sector parameters, 
$m_\DM =10$~kg,
$\vec{x}_{0,\DM} = (0, 2~\text{km}, 2~\text{km})$,
$\vec{v}_{\DM} = (220~\text{km/s}, 0,0)$,
$\lambda = 1$~km,
$\delta_{\text{SM}} \delta_\DM = 1 \times 10^{5}$ and $s = +1$.
The effective coupling constant value chosen is within the limit stated in Eq.~(5) 
from~\cite{Hall:2016usm}.
In this figure we also show the estimated KAGRA noise  in the detuned resonant 
side-band extraction configuration taken from~\cite{KAGRAnoise}
for comparison. We see that the DM signal picks up the characteristic resonance 
frequencies like the suspension thermal noise. We have seen this before also for an 
instantaneous, short-range force; see
Ref.~\cite{Lee:2020dcd} for detailed explanations.

Since we consider here a long-range force which will act on all four suspension 
systems simultaneously we have to extend this formalism to that case. As mentioned 
before in KAGRA the four suspension systems are arranged in an L-shaped 
interferometer and labeled as EX, IX, EY, and IY. The rest positions of the four TM 
mirrors in our lab frame coordinate system are
\begin{align}
\vec{x}_3^{\text{EX}} = (L+l,0,0) \;, \quad
\vec{x}_3^{\text{IX}} = (l,0,0) \;, \quad 
\vec{x}_3^{\text{EY}} = (0,L+l,0) \;, \quad
\vec{x}_3^{\text{IY}} = (0,l,0) \;,
\end{align}
with $l = 25$~m and $L=3000$~m.

Since KAGRA does not measure the absolute position
but a phase difference proportional to the relative positions of the four test mirrors 
we define a horizontal, a vertical, and a combined mode for the full system,
\begin{align}
\tilde{x}_{\text{4m},h}(f) &=[\tilde{x}^{\text{EX}}_{h}(f) - \tilde{x}^{\text{IX}}_{h}(f)]
- [\tilde{x}^{\text{EY}}_{h}(f) - \tilde{x}^{\text{IY}}_{h}(f)], \notag\\ 
\tilde{x}_{\text{4m},v}(f) &=[\tilde{x}^{\text{EX}}_{v}(f) - \tilde{x}^{\text{IX}}_{v}(f)]
- [\tilde{x}^{\text{EY}}_{v}(f) - \tilde{x}^{\text{IY}}_{v}(f)], \notag\\
|\tilde{x}_{\text{4m}}(f)|^{2} &= \text{VHC}^{2} |\tilde{x}_{\text{4m},v}(f)|^{2} + |\tilde{x}_{\text{4m},h}(f)|^{2}.
\label{eq:phasedifference}
\end{align}
Note that the horizontal mode for EX and IX is in $x$-direction while for EY and IY it is 
in $y$-direction. The vertical mode for all four suspension systems is in $z$-direction. 
The calculations for each suspension system 
and mode is in complete analogy to the case discussed above. The observable strain 
amplitude by a DM mass passing by for the full system is then

\begin{equation}
	h_{\text{4m}}(f) = \sqrt{4 f} |\tilde{x}_{\text{4m}}(f)| \;.
\end{equation}

In Fig.~\ref{fig:4mirror} we compare the strain amplitude for the one- suspension setup with the 
four-suspension setup with the same dark sector parameters. We see that there is a difference 
between the DM signal with the four-mirror setup and the one-mirror setup. For example, at  
$f = 10$~Hz $h_{\text{1m}} = 4.17 \times 10^{-21} \; \sqrt{1/\text{Hz}}$
and $h_{\text{4m}} = 1.132  \times 10^{-20} \; \sqrt{1/\text{Hz}}$, which is about a factor three 
difference. The background noise strain at this frequency is equal to 3.86 $\times 10^{-22} \; \sqrt{1/\text{Hz}}$.

That there is such a significant difference between the one- and four-mirror setups should not 
come as a surprise. Here we consider here a long-range Yukawa interaction together with 
Newtonian gravity which implies that there are configurations where there is a significant force 
on more than one of the mirror strings. That is different to our previous work~\cite{Lee:2020dcd} 
where we considered only an instantaneous, very short-range force in one of the mirror 
components. Hence,
we need to consider here the quantity in Eq.~\eqref{eq:phasedifference} and simulate the force
on all mirror components. Please note that the enhancement from one mirror to four mirrors is 
not trivial. As we will see later, the enhancement is not uniform across DM parameters and 
cancellations can happen as well. Hence, the detailed studies done in the following sections 
are necessary.

\begin{figure}
	\centering
	\includegraphics[width=0.9\linewidth]{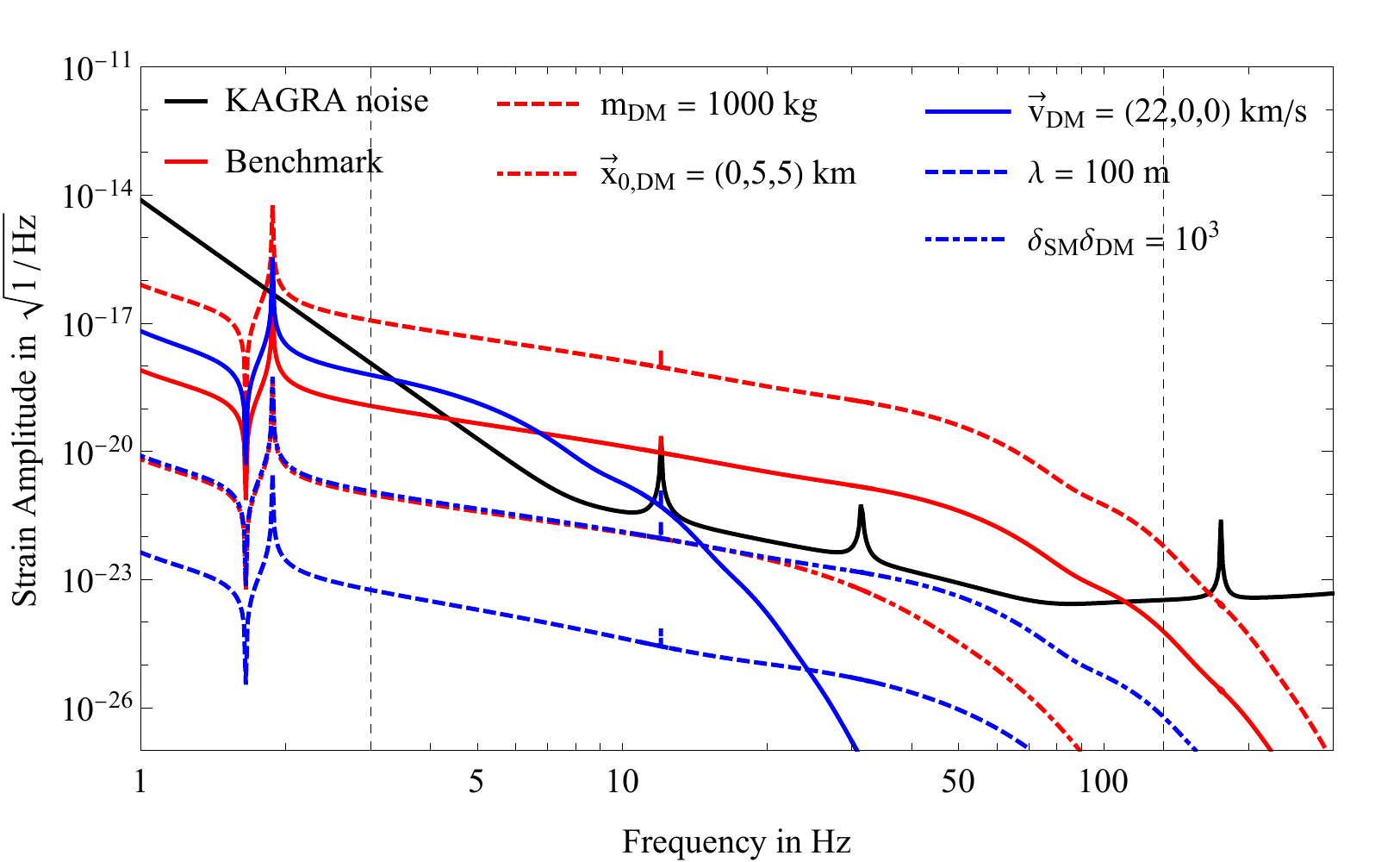}
	\caption{Comparison of DM signals for various dark sector parameters.
		The black line is the expected noise. The benchmark (solid red line) is chosen to have the same parameters with Fig.~\ref{fig:4mirror}, i.e., $m_\DM =10$~kg,
		$\vec{x}_{0,\DM} = (0, 2~\text{km}, 2~\text{km})$,
		$\vec{v}_{\DM} = (220~\text{km/s}, 0,0)$,
		$\lambda = 1$~km,
		$\delta_{\text{SM}} \delta_\DM = 1 \times 10^{5}$ and $s = +1$. The other lines are created using
		the benchmark parameters except the one indicated in the legend.
		The  vertical dashed lines indicate our SNR integration boundaries, $f_\text{min} = 3$~Hz and 133~Hz.}
	\label{fig:strain_estimation}
\end{figure}

We define the signal-to-noise (SNR) for the rest of the paper as
\begin{equation}
	\rho^{2}_{\text{KAGRA}} = \int_{f_{\min}}^{f_{\max}} \diff f \, \frac{4 |\tilde{x}_{\text{4m}}(f)|^{2}}{S_{\text{tot}}(f)}
	\label{eq:snr}
\end{equation}
where we choose $f_{\text{min}}= 3$~Hz and $f_{\text{max}} = 133$~Hz.
Here, we need to make a trade-off. Naively, one can get the largest SNR by integrating 
from zero to infinity which is numerically impossible and we have to choose a finite range. 
At low frequencies one is quickly dominated by seismic noise and it makes sense to cut 
off the integration at $\mathcal{O}(1)$~Hz.
For large frequencies the choice is less obvious. In Fig.~\ref{fig:4mirror}
we see that the DM signal can bend over at a certain frequency and then quickly go to zero. 
But at which frequency the signal starts to bend depends strongly on the chosen parameters,
cf.~Fig.~\ref{fig:strain_estimation} where we chose a number of different strains for some 
plausible dark sector parameters.

At larger frequencies quantum noise dominates apart from some resonances from the 
thermal noise which we neglect for now. In KAGRA the quantum noise has roughly a 
V-like shape and we can split it into two frequency ranges. Between about 10~Hz and 
$f_{\text{int}} \approx 93$~Hz
\begin{align}
 S_{\text{tot}}(f) \approx \alpha \frac{f_{\text{int}}^4}{f^4} \;,
\end{align} 
and for frequencies larger than $f_{\text{int}}$
\begin{align}
 S_{\text{tot}}(f) \approx \alpha \frac{f_{\text{int}}^2}{f^2} \;,
\end{align}
where we neglected the resonance peaks from the thermal noise.

In many cases the DM signal on the other hand scales roughly as
\begin{align}
 |\tilde{x}_{\text{4m}}(f)|^{2} \approx \beta \frac{f_{\text{int}}^3}{f^3} \;,
\end{align}
where we again neglect (anti-)resonances and also ignore that at some
frequency the signal can drop much faster.

Then
\begin{align}
 \rho^{2}_{\text{KAGRA}} &= \int_{f_{\min}}^{f_{\max}} \diff f \, \frac{4 |\tilde{x}_{\text{4m}}(f)|^{2}}{S_{\text{tot}}(f)} 
 \approx  \frac{4 \, \beta }{\alpha} \int_{f_{\min}}^{f_{\text{int}}} \diff f \,   \frac{f}{f_{\text{int}}} 
  + \frac{4 \, \beta }{\alpha} \int_{ f_{ \text{int} } }^{f_{\max}} \diff f \,   \frac{f_{\text{int}}^5}{f^5} \nonumber\\
  &= \frac{4 \, \beta }{\alpha} \left[ \frac{f_{ \text{int} }}{2} \left(  1 - \frac{f_{\min}}{f_{ \text{int} }} \right)  + \frac{f_{ \text{int}}}{4} \left( \frac{f_{ \text{int}}^4}{f_{\max}^4} -1 \right) \right] \nonumber\\
  &= \frac{4 \, \beta \,  f_{ \text{int} } }{\alpha} \left[ \frac{1}{4} - \frac{f_{\min}}{2 \, f_{ \text{int} }}  +  \frac{f_{ \text{int}}^4}{4 \, f_{\max}^4}  \right]  \;.
\end{align}
What we see from this approximation is that the frequency range below $f_{\text{int}}$ is 
much more important. Going to higher frequencies also helps but the contributions are 
comparatively suppressed. And this is under the assumption that the signal falls slowly 
even for large frequencies which is often not the case.

In fact, in our analysis later we rarely care about the precise value of
$ \rho^{2}_{\text{KAGRA}}$. We are mostly interested in relative and qualitative
features and the region where $ \rho^{2}_{\text{KAGRA}}  > 1$. This
explains why we fix $f_{\max} = 133$~Hz which is quite a bit larger
than $f_{\text{int}}$ to be on the safe side while keeping computation time feasible.
We will also show that, for instance,
the expected observable event rate at KAGRA depends only weakly on
$f_{\max}$ for that large frequency later on when we discuss our numerical results. 
Nevertheless, we want to stress that in the following sections we will simulate the full SNR 
including the resonances and not use the above approximations. They are used only to 
justify the chosen frequency range for the numerical integration.

\section{DIRECTIONAL AND PARAMETER DEPENDENCE OF THE SNR}
\label{sec:Directions}

In this section we want to have a closer look at the directional and dark sector parameter 
dependence of the SNR which we just defined. Since KAGRA measures the relative 
displacement of the detectors in two directions there is some apparent intrinsic directionality 
related to the initial position (the closest point to the mirrors) and velocity of the passing DM 
particle.

It is challenging to discuss the possible six-dimensional DM phase comprehensively. Therefore 
we will first fix the initial position and just change the direction of the DM velocity and then 
change the initial position and consider a velocity orthogonal to the initial position vector to give 
some impression of some of the important features.

\begin{figure}
	\centering
	\includegraphics[width=0.9\linewidth]{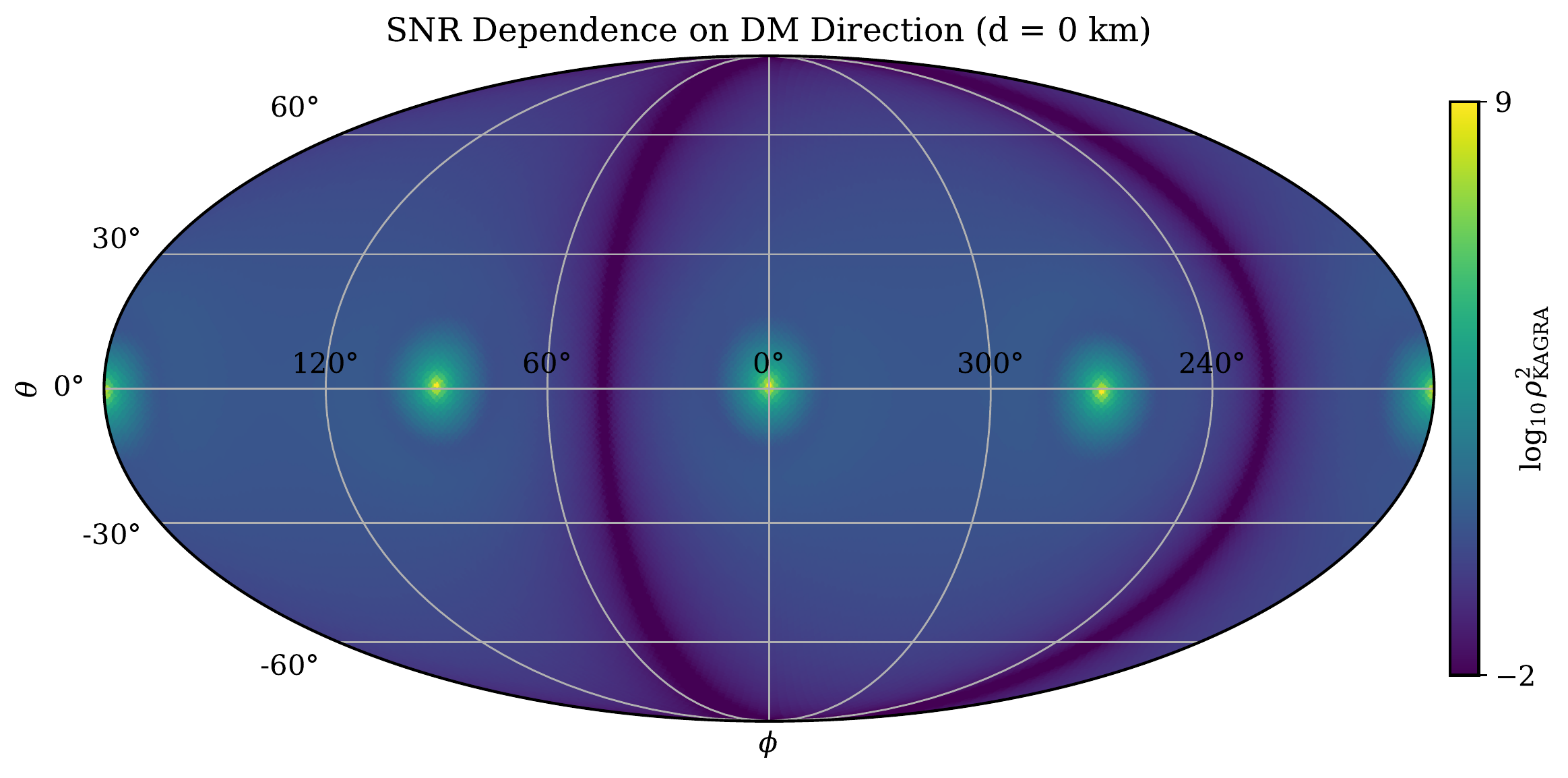}
	\caption{
		Dependence of the SNR on the direction of the DM velocity. For the initial position
		we chose the origin and we fix the DM velocity to 220~km/s. The angles of this
		Mollweide projection fix the direction of the velocity. For the other dark sector parameters
		we chose $m_{\DM} = 10$~kg, $\lambda = 1$~km, and $\delta_{\text{SM}} \delta_\DM = 10$.
		We also implemented a cutoff to the color scale to keep all features visible.
		For further explanations, see main text.
		}
	\label{fig:Mollweide1}
\end{figure}

For the first case, we set $\vec{x}_{0,\DM}$ to be the origin of the coordinate system which 
is at the intersection of the two interferometer arms. We also fix the absolute DM velocity to be 
$v_0 \equiv |\vec{v}_{\DM}| = 220$~km/s and then only vary its direction. Hence, we are left
with two parameters (angles) and we set
\begin{equation}
\vec{v}_{\text{DM}} = v_0 (\cos \phi \, \cos \theta,  \sin \phi \, \cos \theta, \sin \theta) \;.
\end{equation}
We then scan $\log_{10} \SNR$ over the polar angle from $90^\circ$ to $-90^\circ$
and the azimuthal angle from $0^\circ$ to $360^\circ$ and plot the result in a Mollweide projection
in Fig.~\ref{fig:Mollweide1}. For the other dark sector parameters we chose $m_{\DM} = 10$~kg,
$\lambda = 1$~km and $\delta_{\text{SM}} \delta_\DM= 10$.

The four brightest spots around $\theta = 0^\circ$, which are $\phi = 0^\circ$, $90^\circ$, 
$180^\circ$ and $360^\circ$, correspond to when the DM particle is basically going along the 
$x$- and $y$-direction,
that means it is getting very close to the suspension systems and the exerted force is large. 
The force and hence the SNR is actually getting so large that we took the logarithm of the SNR 
in the figure and restricted the color scale to keep all relevant features visible.

Apart from the four very bright spots there is another feature. If the DM
passes by in the direction for $\phi \approx 45^\circ$ and $225^\circ$ the SNR is suppressed. 
That is when the DM passes through the right angle spanned by the L. In this case the force 
exerted on both arms is very similar and the effect cancels out when taking the difference 
between the two arms, cf.~Eqs.~\eqref{eq:phasedifference}.
On the other hand, for the direction orthogonal to that case, the effect adds up when taking the 
difference and the signal is enhanced. What we can also see is the mirror symmetry about the 
$x$-$y$-plane leading to an intrinsic ambiguity if one would try to reconstruct the DM velocity.

\begin{figure}
	\centering
	\includegraphics[width=0.9\linewidth]{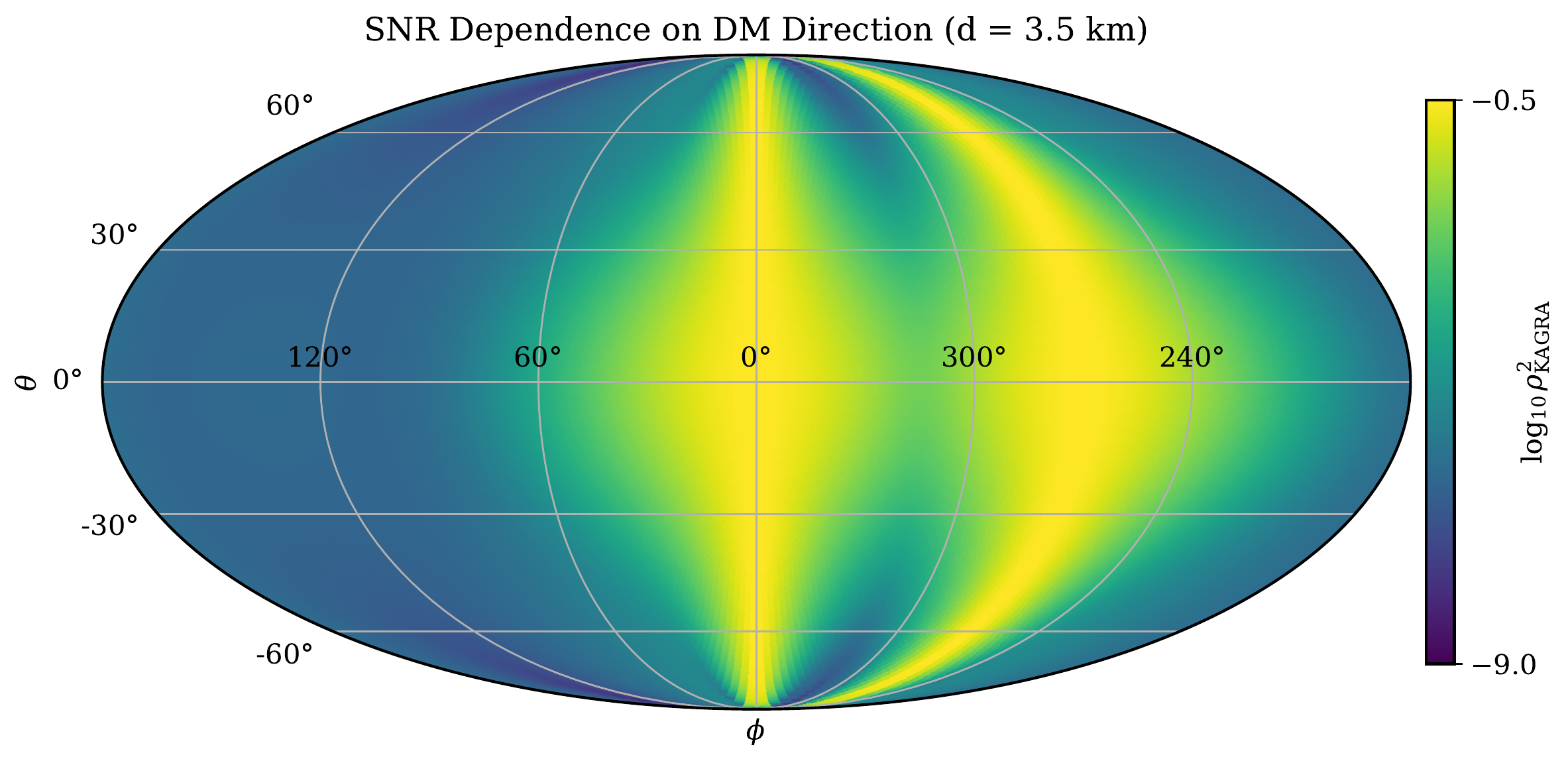}
	\caption{
		Dependence of the SNR on the initial position and the direction of the DM velocity. We
		first fixed the DM velocity to 220~km/s and its direction according to the angles in this Mollweide
		projection and then we choose the initial position to be orthogonal to it.
		For the other dark sector parameters
		we choose $m_{\DM} = 10$~kg, $\lambda = 1$~km, and $\delta_{\text{SM}} \delta_\DM = 10$. We also implemented a cutoff to the color scale to keep all features visible.
		For further explanations, see main text.
		}
	\label{fig:Mollweide2}
\end{figure}

\begin{figure}
	\centering
	\includegraphics[width=0.45\linewidth]{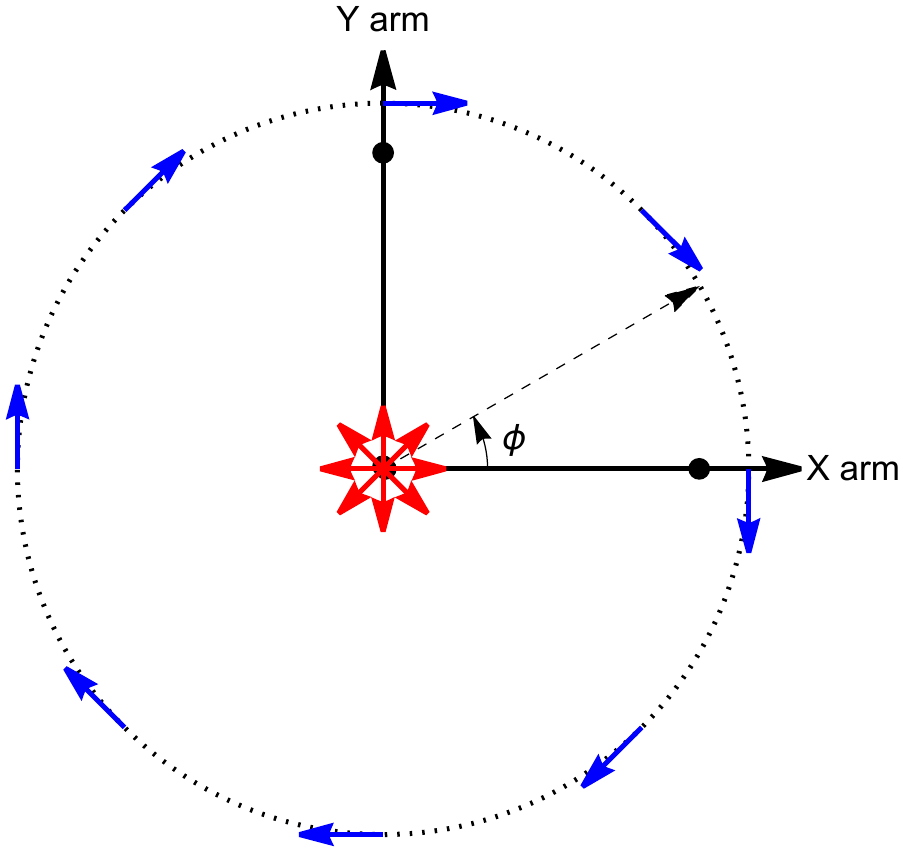}
	\caption{
		Sketch to clarify the Mollweide projections in Figs.~\ref{fig:Mollweide1}
		and \ref{fig:Mollweide2}. This sketch is a cut in the $x$-$y$ plane ($\theta = 0^\circ$) with the
		KAGRA detectors along the interferometer arms labeled with a black dot.
		The red (blue) arrows correspond to the DM velocity vectors considered
		in Fig.~\ref{fig:Mollweide1} (\ref{fig:Mollweide2}). The dashed circle is the imaginary
		line through which the DM is passing in Fig.~\ref{fig:Mollweide2}. In Fig.~\ref{fig:Mollweide1}
		DM is going through the origin, the center of the circle. For further explanations,
		see main text.
		}
	\label{fig:MollweideSketch}
\end{figure}

In the second case in Fig.~\ref{fig:Mollweide2}, we fix the dark sector parameters as before in 
Fig.~\ref{fig:Mollweide1} where 
$m_{\DM} = 10$~kg, $\lambda = 1$~km and $\delta_{\text{SM}} \delta_\DM = 10$.
We also fix the DM velocity to be 220~km/s and the Mollweide projection maps its different 
directions as before. That means the two angles again have the
same meaning. What changes now is the initial position. Here the DM does not pass through 
the origin but it passes through a ring in the $x$-$y$ plane with a radius of 3.5~km. The initial 
position vector is chosen orthogonal to the velocity vector which still leaves a sign ambiguity. 
This is resolved here such that when $\vec{v}_{\text{DM}}$ was pointing in negative $y$-direction,
i.e., $\vec{v}_{\text{DM}} = (0,- |\vec{v}_{\text{DM}}|,0)$, the position vector was given
by $\vec{x}_{0,\text{DM}} = (\text{+3.5 km},0,0)$. Similarly for 
$\vec{v}_{\text{DM}} = (|\vec{v}_{\text{DM}}|,0,0)$
the position vector was given by $\vec{x}_{0,\text{DM}} = (0,\text{+3.5 km},0)$. 
We have made a sketch in Fig.~\ref{fig:MollweideSketch} to make this clearer.
That explains why these two
directions are brighter (have a larger SNR) in the Mollweide projection in Fig.~\ref{fig:Mollweide2} 
since here the DM comes closest to one of the suspension systems. Note that the angle $\phi$ 
labels the direction of the DM velocity. And in this figure
$\phi = 0^\circ$ means it passes by close to the outer detector of the $Y$ arm
and for $\phi = 270^\circ$ it passes by close to the outer detector of the $X$ arm.If the DM 
velocity points in the opposite direction it also passes by at the other side of the circle 
and it is much further away from any suspension system. And hence, the effect on KAGRA and 
the SNR are smaller.

At this point we can also already study how the boundary $\rho_{\text{KAGRA}}^2 = 1$ changes
when changing the dark sector parameters. This boundary is important since it will be used in the
next section to classify which events are considered observable.

\begin{figure}
	\centering
	\includegraphics[width=0.46\textwidth]{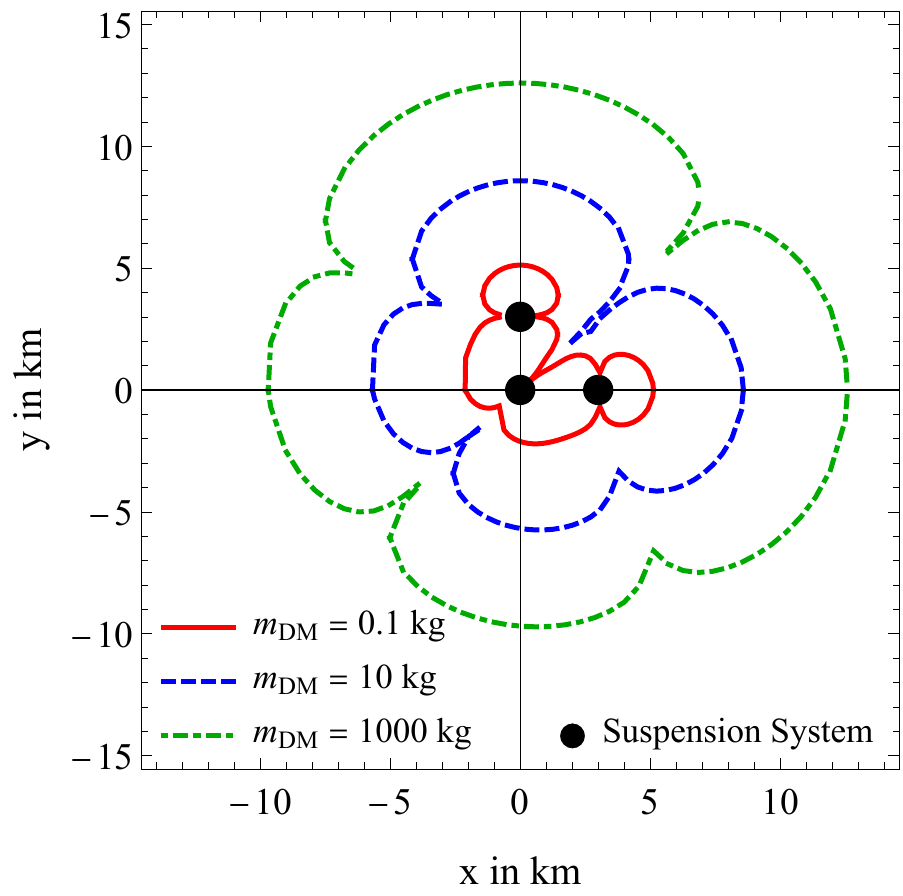} \hfill
	\includegraphics[width=0.46\textwidth]{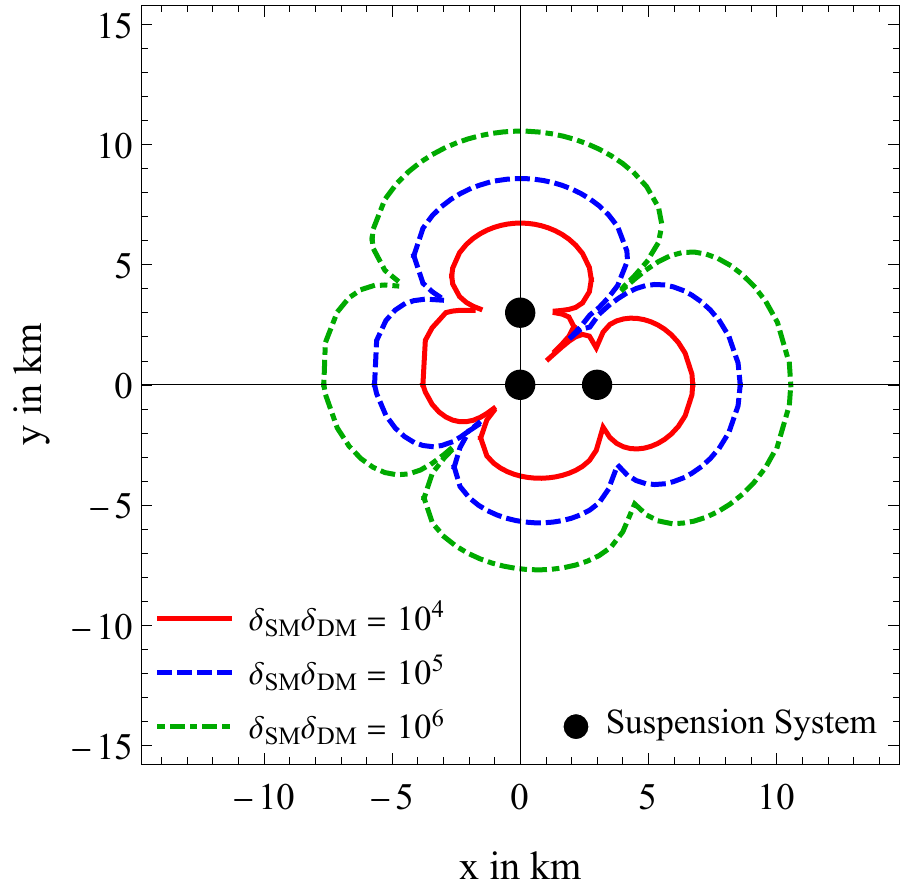}
	\includegraphics[width=0.46\textwidth]{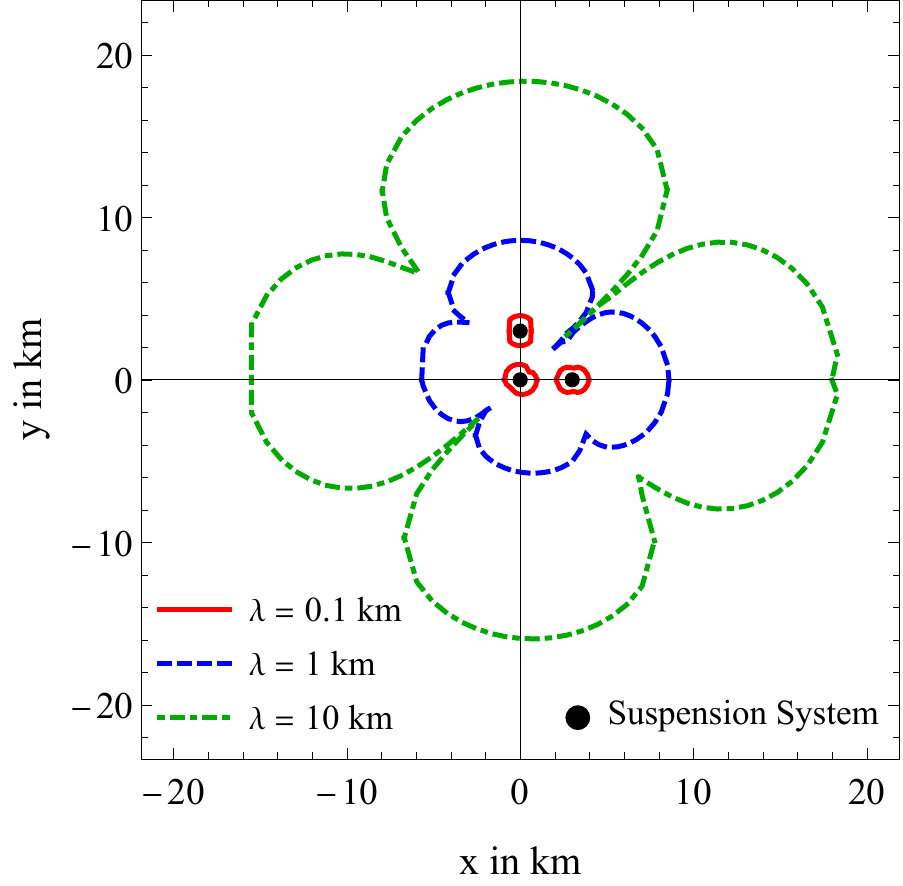}
	\caption{
		The dependence of the boundary $\rho_\text{KAGRA}^2 = 1$ in the
		$x$-$y$-plane on the dark sector parameters. Inside the contours $\rho_\text{KAGRA}^2 > 1$.
		We have chosen the DM
		velocity to be 220~km/s in $z$-direction. In the upper left plot
		$\delta_{\text{SM}} \delta_\DM = 10^{5}$, $\lambda = 1$~km,
		in the upper right plot
		$m_\text{DM} = 10$~kg, $\lambda = 1$~km,
		and in the lower plot
		$m_\text{DM} = 10$ kg, $\delta_{\text{SM}} \delta_\DM = 10^{5}$.
	}
	\label{fig:boundary}
\end{figure}

In Fig.~\ref{fig:boundary} we assume the DM velocity to be in $z$-direction with a 
value of 220~km/s. The lines in the plots then depict the position where the DM 
passes the $x$-$y$-plane and where $\rho_\text{KAGRA}^2 = 1$ for the chosen 
parameters. The positions of the suspension systems are depicted as black dots 
and the inner dots practically overlap.

In all three plots the dents in the lines are the cancellations which we have already 
partially seen before in Fig.~\ref{fig:Mollweide1}. Apart from that we can also clearly 
see the forces on the mirror and hence the SNR scales. As the DM mass, 
$\delta_\text{SM}\delta_\text{DM}$ and $\lambda$ get larger,
the forces felt by the suspension systems become larger as well. Hence, the area where
$\rho_\text{KAGRA}^2 > 1$ becomes larger. For example the ratio between the area for
$m_\text{DM} = 1000$ kg and $m_\text{DM} = 10$ kg is around 2.57 while the ratio
for $m_\text{DM} = 10$ kg and $m_\text{DM} = 0.1$ kg is around 4.75.
While the mass scales by a factor of 100, the area grows only by an order one factor. So 
we see that the scaling is nonlinear and that it does not even follow a simple power law. 
Since the number density of the dark matter is inversely proportional to the mass of the 
dark matter, this observation will have an impact on the sensitivity as discussed in the next 
section.

In the cases of changing  $\delta_\text{SM}\delta_\text{DM}$ and $\lambda$,
the area for $\rho_\text{KAGRA}^2 > 1$ grows larger without any reduction on the DM density. 
Hence, we expect that the observable rate will grow larger as any of the parameters increase. 
In the case of varying $\delta_\text{SM}\delta_\text{DM}$, we found the area ratio to be 
$A_{10^6}/A_{10^5} = 1.68$ and $A_{10^5}/A_{10^4} = 1.98$ where the subscripts denote the 
values of  $\delta_\text{SM}\delta_\text{DM}$. For the $\lambda$ dependence there is another 
remarkable feature. For a small value of $\lambda$, the DM interacts significantly with one mirror 
only as shown in the case of $\lambda = 0.1$~km. Hence we found that while the area ratio 
$A_{10}/A_{1} = 5.17$, the ratio $A_{1}/A_{0.1}$ becomes 50.4. Here the subscripts denote the 
values of $\lambda$ in km.

Finally we note that in all the plots in this section we use the dark matter velocity to be 220~km/s, 
which is a typical value in the standard halo model. For different DM velocities, the shapes and 
numbers might change in some plots. However the overall trends do not change so we will not 
discuss these changes in detail here. Besides, we used a much lower effective coupling constant 
value when plotting the Mollweide projections to make sure that the numerical values of the SNR 
do not blow up when the DM particles pass through along the beamline. With that we can present 
a much clearer picture of the SNR variation. Similar is the case for the DM velocity, where if the 
coupling constant value is changed, the overall outcome will remain the same. In the later sections, 
we will scan over the coupling constant when we calculate the event rate.

\section{EXPECTED EVENT RATES}
\label{sec:Rates}

To provide potential discovery prospects of KAGRA for the type of DM we consider here that we 
have to calculate the expected observable event rate, $R_{\text{obs}}$. That is the number of 
expected events in a unit time with an SNR value greater than a threshold value we will set to one.

To derive this rate we begin with the distribution of SNR values which is given by
\begin{equation}
	g(\rho^{2})= \frac{1}{\mathcal{N}_g}  \iint f_v(\vec{v}_{\DM}) 
	\, f_x(\vec{x}_{0,\DM}) \, \delta(\rho^{2}_{\text{KAGRA}}(\vec{v}_{\DM},\vec{x}_{0,\DM})-\rho^{2}) \, 
	\diff^3 \vec{v}_{\DM} \, \diff^3 \vec{x}_{0,\DM}
\end{equation}
where $\mathcal{N}_g = \int_{0}^{+\infty} g(\rho^{2}) \, \diff \rho^{2}$ is a normalization 
factor and $\vec{v}_{\DM}$ and $\vec{x}_{0,\DM}$ are 
the three-dimensional velocity and initial position of the DM particles, cf.,~Sec.~\ref{sec:Framework}.
The functions $f_v$ and $f_x$ are normalized distribution functions for velocity and position.
This equation is difficult to evaluate since it is nontrivial to find
the initial conditions for a given value of $\rho^2$. In other words it is difficult to find
the phase space values where the Dirac $\delta$-function does not vanish.

But since we are only interested in the number of events with an SNR value greater than
a certain threshold value, $\rho_{\text{th}}^{2}$, we can calculate instead
\begin{align}
	p(\rho_{\text{th}}^{2}) &= \int_{\rho_{\text{th}}^{2}}^{\infty} g(\rho^{2}) \, \diff \rho^{2} \nonumber\\
	&= \frac{1}{\mathcal{N}_p}  \int f_v(\vec{v}_{\DM}) 
	\, f_x(\vec{x}_{0,\DM}) \, \theta(\rho^{2}_{\text{KAGRA}}(\vec{v}_{\DM},\vec{x}_{0,\DM})-\rho_{\text{th}}^{2}) \, 
	\diff^3 \vec{v}_{\DM} \, \diff^3 \vec{x}_{0,\DM} \;,
\end{align}
which is much easier to calculate than $g(\rho^{2})$. The $p(\rho_{\text{th}}^{2})$  is the fraction 
of DM events which would give an SNR greater than $\rho_{\text{th}}^{2}$.
Since the total fraction of events with $\rho^{2} > \rho_{\text{th}}^{2}$ and  $\rho^{2} \leq \rho_{\text{th}}^{2}$
must be one we can also easily calculate, the normalization factor 
\begin{align}
\mathcal{N}_p &= \int_{\rho_{\text{th}}^{2}}^{\infty} g(\rho^{2}) \, \diff \rho^{2} + \int_0^{\rho_{\text{th}}^{2}} g(\rho^{2}) \, \diff \rho^{2} \nonumber\\
 &= \int f_v(\vec{v}_{\DM}) 
\, f_x(\vec{x}_{0,\DM}) \, 
\diff^3 \vec{v}_{\DM} \, \diff^3 \vec{x}_{0,\DM} = 1 \,.
\end{align}

We assume that all initial positions are equally likely, that means $f_x = 1/V$ where $V$ is 
the considered volume. Naively one might want to take the solar system as the volume or 
something of similar size. But this is not necessary since the volume where the DM can 
have a significant impact on the detector is much smaller and using a smaller volume is 
numerically more efficient. To estimate the maximum radius of positions we have to 
consider to get a detectable signal we define $F_\text{min}$ as the minimal force on the 
KAGRA mirrors which could be resolved experimentally. From
\begin{equation}
	F_\text{min} = |\vec{F}_{\text{DM}}(r_{\text{max}})|
	\label{eq:Fmin}
\end{equation}
we can then calculate a maximum distance between the DM particle and the KAGRA 
mirrors where the DM might still be detected. For DM to pass by in a further distance 
the force would be weaker and a detection would be impossible.

We can estimate
\begin{equation} \label{eq:fmin}
 F_{\text{min}} \approx \frac{m_3 \, g \, \Delta d_\text{min}}{ \, l} \;,
\end{equation}
where $g$ is the gravitational acceleration and $l$ = 0.35 m is the length of the wire
that connects the blade springs and the test mirror with mass $m_3$. The minimum 
displacement that can be resolved is given by 
$\Delta d_\text{min} = h_\text{min} \, L \, \sqrt{\Delta f}$ where we use the strain 
sensitivity of KAGRA $h_\text{min} = 2 \times 10^{-24} \text{ Hz}^{-1/2}$
at $f = 100$~Hz~\cite{KAGRA:2018plz} and $L$
is the interferometer arm length. The bandwidth $\Delta f$ is taken as 10~Hz
which is a typical value used for analysis in Gravitational Wave (GW) 
detectors~\cite{LIGOScientific:2016emj}.
With these numbers, we get $F_{\text{min}} \approx 1.21 \times 10^{-17}$~kg~m/s$^{2}$.

Putting this value back into Eq.~\eqref{eq:Fmin}, we can then calculate the corresponding maximum
distance between the DM particle and one of the KAGRA 
mirrors for a given set of dark sector parameters which we need to consider in our scans.
For instance, for $\lambda = 1$~km, $\delta_{\text{SM}} \delta_\DM = 5 \times 10^{5}$ and 
$M_{\DM} = 10$~kg we find $r_{\text{max}} = 15.8$~km.
But this is the distance from one of the mirrors. To take into account all suspension systems,
and as a simple and conservative choice, we consider as volume $V$ a sphere
with radius
\begin{equation}
	R_{\text{max}} = 1.1 \times (r_{\text{max}} + L) 
\end{equation}
centered on the origin of our lab frame.
We checked numerically that this estimate is quite robust and could not find any parameter 
points where we should have chosen a larger radius. In fact, in many cases this radius is 
significantly overestimated but since it is simple to implement we use this volume in our 
numerical calculation.

Apart from that volume restriction we used another observation in our calculations to scan 
the initial conditions for DM more efficiently. In our analysis we ignore the backreaction of the 
detector on the DM particle. We assume it follows a straight line with a constant velocity to a 
good approximation. That implies we get the same SNR value no matter what point on the 
line we choose as initial position.

The SNR depends just on the absolute value
$|\tilde{x}_{\text{4m}}(f)|$. Choosing different initial points on a DM trajectory would lead
to a different global phase of $\tilde{x}_{\text{4m}}(f)$ which disappears when taking
the absolute value. We can therefore represent each DM trajectory by any point on it and 
we choose the point closest to the origin of our coordinate system.

This point has the advantage that the position vector
and the DM velocity are orthogonal to each other
\begin{align}
\hat{v}_{\DM} \cdot \vec{x}_{0,\DM} = 0 \;,
\end{align}
where we introduce the normalized DM velocity vector, $\hat{v}_{\DM}$, for later convenience.
In fact, this defines a plane, not just a trajectory but our argument remains true for
every trajectory on that plane as well. Note that the SNR value is in general different
for different directions on that plane due to the directional sensitivity of the detector.
The length of any trajectory within the considered sphere is
$2 \sqrt{R_{\text{max}}^2 - \vec{x}_{0,\DM}^2}$.

We can then introduce this insights into our integral as a $\delta$-function
multiplied with the length of the trajectory
\begin{align}\label{eq:prhoth}
p(\rho_{\text{th}}^{2})
&= \frac{3}{2 \, \pi R_{\text{max}}^3 } \int f_v(\vec{v}_{\DM}) 
\, \theta(\rho^{2}_{\text{KAGRA}}(\vec{v}_{\DM},\vec{x}_{0,\DM})-\rho_{\text{th}}^{2}) \theta( R_{\text{max}} - |\vec{x}_{0,\DM}| )  \times \nonumber\\
 &\phantom{= \frac{3}{4 \pi R_{\text{max}}^3 } \int}  \sqrt{R_{\text{max}}^2 - \vec{x}_{0,\DM}^2} \, \delta(\hat{v}_{0,\DM} \cdot \vec{x}_{0,\DM})  \, 
\diff^3 \vec{v}_{\DM} \, \diff^3 \vec{x}_{0,\DM} \nonumber\\
&= \frac{3}{2 \, \pi R_{\text{max}}^3 }  \int \frac{|\vec{v}_{\DM}| f_v(\vec{v}_{\DM}) }{|v_{z,\DM}|} \sqrt{R_{\text{max}}^2 - (\vec{x}'_{0,\DM})^2} \; \theta( R_{\text{max}} - |\vec{x}'_{0,\DM}| )  \times \nonumber\\
&\phantom{= \frac{3}{4 \pi R_{\text{max}}^3 } \int}  \theta(\rho^{2}_{\text{KAGRA}}(\vec{v}_{\DM},\vec{x}'_{0,\DM})-\rho_{\text{th}}^{2})  \,
\diff^3 \vec{v}_{\DM} \, \diff x_{0,\DM} \diff y_{0,\DM} \;,
\end{align}
where
\begin{equation}
\vec{x}'_{0,\DM} = \left(x_{0,\DM}, y_{0,\DM}, - \frac{ x_{0,\DM}  v_{x,\DM} + y_{0,\DM} v_{y,\DM} }{v_{z,\DM}} \right) \;.
\end{equation}

These restrictions do not only limit the coordinate integral to a meaningful volume and get rid of 
one integration; they have another useful side effect. For the Fourier integral to get 
$\tilde{x}_{4\text{m}}$ we only have to integrate over the time interval from $-T$ to $T$ where 
$T = R_{\text{max}}/|\vec{v}_{\DM}|$ since only during that time the DM could have a 
measurable impact on the detector.

What is left to discuss before we can evaluate the integral is the DM velocity distribution.
In this work, we employ for it the Standard Halo Model (SHM) given by
a Maxwellian distribution in the galactic rest frame. The DM velocity in this frame is labeled
as $\vec{v}^{\, \text{gal}}_{\DM}$. Its distribution has a dispersion
$\sigma_{v}$ and is truncated at the escape velocity $v_{\text{esc}}$~\cite{Spergel:1987kx,Gondolo:2002np,Bozorgnia:2011tk}. But we are working here in the
lab frame. The relationship between the DM velocity in galactic rest frame and the lab frame 
is given by
\begin{equation}
 \vec{v}^{\, \text{lab}}_{\DM} = \vec{v}^{\, \text{gal}}_{\DM} + \vec{v}_{\text{lab}}(t) \;,
\end{equation}
where $\vec{v}_{\text{lab}}(t)$ is the time-dependent velocity of the lab in the galactic rest frame.
The time-dependence here leads to the well-known daily and annually modulation of the 
expected DM rate, cf.~\cite{Spergel:1987kx,Gondolo:2002np,Bozorgnia:2011tk}.

The DM velocity distribution in the lab frame is then given as
\begin{equation} \label{eq:dist}
	f_v(\vec{v}^{\, \text{lab}}_{\DM}, t)  \equiv f_\mathrm{SHM}(\vec{v}^{\, \text{lab}}_{\DM}(t)) = \frac{1}{(2\pi \sigma_v^2)^{3/2}N_\text{esc}} \, 
	\exp \left( - \frac{|\vec{v}^{\, \text{gal}}_{\DM} + \vec{v}_{\, \text{lab}}(t)|^2}{2\sigma_v^2}\right) \\
\end{equation}
for $|\vec{v}^{\, \text{gal}}_{\DM}| < v_{\text{esc}}$ and zero otherwise.
The normalization constant is given by
\begin{equation}
	N_\text{esc} = \erf \left( \frac{v_\text{esc}}{\sqrt{2}\sigma_v}\right) - \sqrt{\frac{2}{\pi}} 
	\frac{v_\text{esc}}{\sigma_v} \exp \left( -\frac{v_\text{esc}^2}{2\sigma_v^2}   \right)\,.
\end{equation}

In our numerical calculations we performed the velocity integration actually over
$\vec{v}^{\, \text{gal}}_{\DM}$ which is trivial since we fixed $\vec{v}_{\, \text{lab}}(t)$
to be a constant value. For notational simplicity, we will remain to use $\vec{v}_{\DM}$ as 
the DM velocity in lab frame from now on even in integrals. 

\begin{figure}
	\centering
	\includegraphics[width=0.48\textwidth]{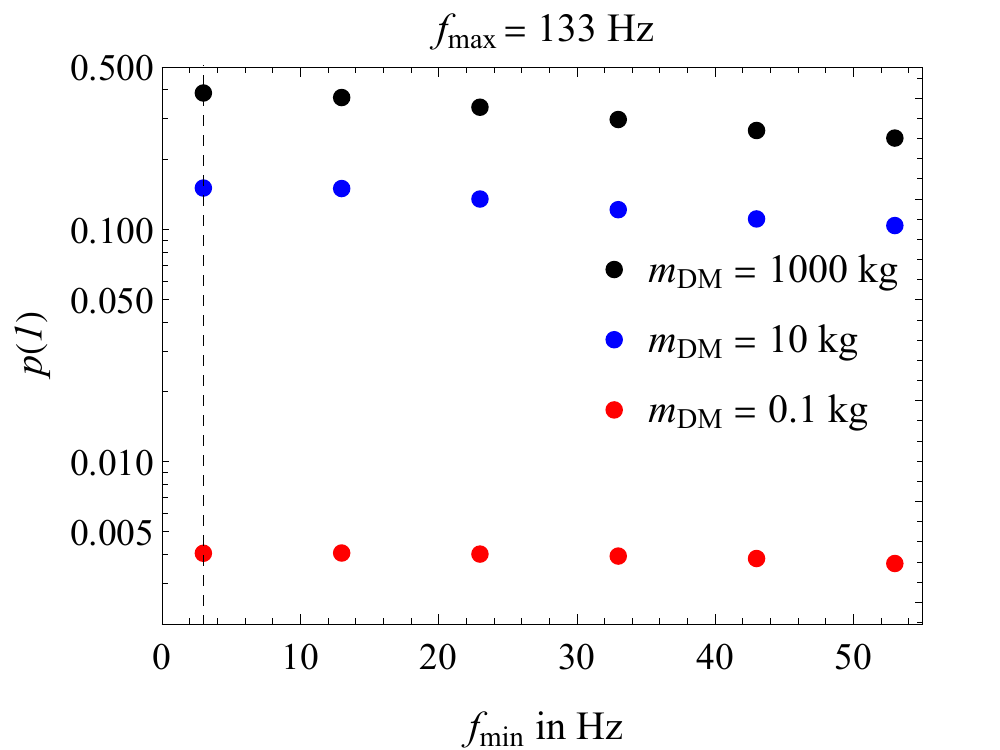} \hfill
	\includegraphics[width=0.48\textwidth]{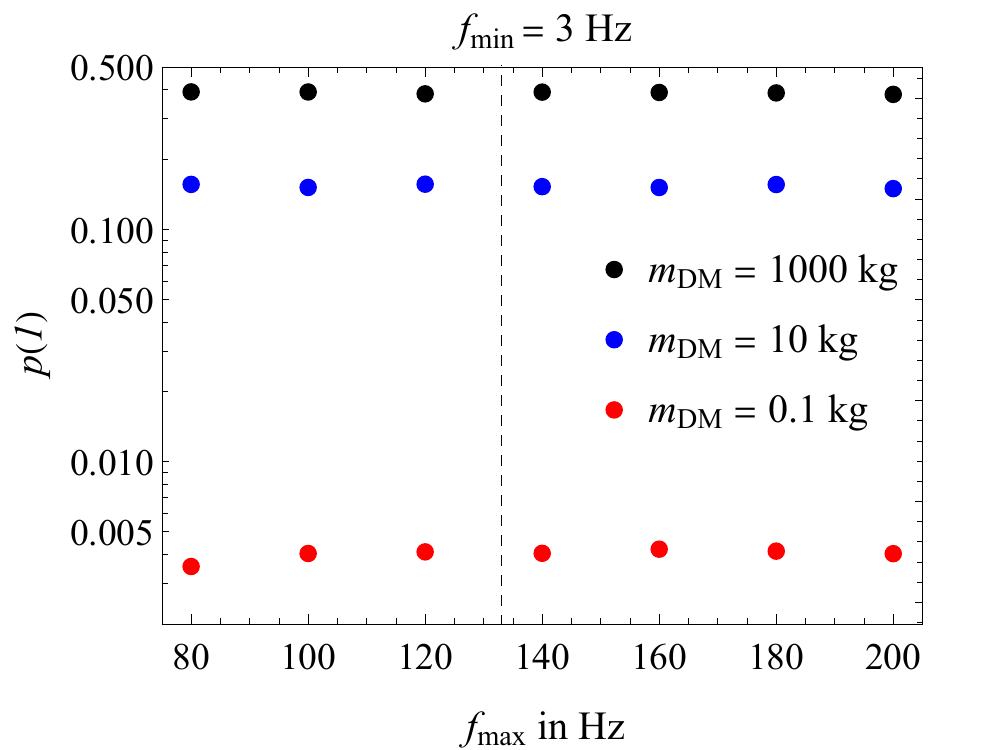}
	\caption{
		The fraction of DM events with $\rho_{\text{th}}^{2} \geq 1$
		as a function of $f_\text{min}$ (on the left) and $f_\text{max}$ (on the right).
		The red points correspond to $m_\text{DM} = 0.1$~kg,
		$\delta_{\text{SM}} \delta_\DM = 10^{4}$, $\lambda = 0.1$~km,
		while the blue points correspond to $m_\text{DM} = 10$ kg,
		$\delta_{\text{SM}} \delta_\DM = 10^5$, $\lambda = 1$ km, and the black points correspond 
		to $m_\text{DM} = 1000$ kg, $\delta_{\text{SM}} \delta_\DM = 10^5$, $\lambda = 1$ km.
		We label $f_\text{min} = 3$~Hz and $f_\text{max} = 133$ with vertical dashed lines.}
	\label{fig:pDependence}
\end{figure}

At this point it is important to point out that this assumed DM velocity distribution should be 
taken with a grain of salt. Not only that this velocity distribution might be incorrect within our 
astronomical neighborhood, the gravitational and Yukawa backreaction and other new forces 
might modify this via friction, attenuation, focusing, clustering, etc. This could lead to small 
deviations but it could even imply that a significant fraction of DM could not penetrate the 
Earth to reach KAGRA.

Additional new interactions are highly model dependent and a discussion of them is beyond 
the scope of the current work. Considering only the Yukawa interaction, the velocity 
distribution can still be modified. In~\cite{Baum:2022duc} they considered a similar scenario 
but with a much larger Yukawa range $\lambda \sim \mathcal{O}$(AU) and they found 
deviations induced by 
interactions with the sun. In our case due to the much shorter range we expect deviations 
mainly induced by the Earth itself. This can lead to an enhancement of the expected signal 
rates due to local clustering or a depletion due to shielding and repulsion effects. Since our 
work here focuses on the modeling of DM interactions in KAGRA we leave the modeling of 
a more realistic velocity distribution and number density including possible new interactions 
for a future work.

Now we have everything together to numerically calculate $p(\rho_{\text{th}}^{2})$ for any 
given value of dark sector parameters using the Vegas Monte Carlo routine in 
Python~\cite{Lepage:2020tgj}.
As a warmup we want to pick up on a previous issue and study how
$p(\rho_{\text{th}}^{2})$ depends on the integration boundaries for the SNR, 
$f_{\text{min}}$ and $f_{\text{max}}$, cf.~Eq.~\eqref{eq:snr}.
In Fig.~\ref{fig:pDependence} we show the dependence of $p(\rho_{\text{th}}^{2} = 1)$ for 
these two boundaries independently leaving the other boundary fixed to our chosen value.
We can see that $p(\rho_{\text{th}}^{2} = 1)$ 
depends only weakly on $f_{\max}$
as long as it is above $f_{\text{int}}$ justifying our choice of $f_{\max} = 133$~Hz.
We also see that the difference for $f_{\text{min}}$ of $\mathcal{O}(1)$ is marginal.

We now turn finally to the expected observable event rate
\begin{equation}
  R_{\text{obs}} = p(\rho_{\text{th}}^{2})  \, R_{\text{total}}  \;,
\end{equation}
where $R_{\text{total}}$ is the rate of DM particles entering
(or leaving) the sphere given by the radius $R_{\text{max}}$ in a given time
and $\rho_{\text{th}}^{2}$ is the threshold value for observability.

To calculate it though we need to first calculate the total rate of DM particles passing by.
That is the number flux of
DM particles $\vec{\Phi}_{\text{DM}}$ going through the surface of the sphere,
i.e., the integral
\begin{align}
	R_{\text{total}} &= \frac{1}{2} \int |\vec{\Phi}_{\text{DM}} \cdot \vec{S}| f_v(\vec{v}_{\text{DM}}) \diff^2 \vec{S} \diff^3 \vec{v}_{\text{DM}} \;,
\end{align}
where the factor $1/2$ is to avoid double counting (we only count DM particles
entering OR leaving the sphere) and $\vec{S}$ is the normalized normal vector on the surface 
of the sphere.
Now we evaluate this formula
\begin{align}
	R_{\text{total}} &= \frac{1}{2} \int |\vec{\Phi}_{\text{DM}} \cdot \vec{S}| f_v(\vec{v}_{\text{DM}}) \diff^2 \vec{S} \diff^3 \vec{v}_{\text{DM}}
	= \frac{  n_{\text{DM}} R_{\text{max}}^2  }{2}  \int |\vec{v}_{\text{DM}}\cdot \vec{e}_{r,S} | f_v (\vec{v}_{\text{DM}}) \diff^2 \vec{S} \diff^3 \vec{v}_{\text{DM}} \nonumber\\
	&= \frac{  n_{\text{DM}} R_{\text{max}}^2  }{2} \int  |\vec{v}_{\text{DM}}|^3 | \vec{e}_{r,\text{DM}} \cdot \vec{e}_{r,S} |  f_v(\vec{v}_{\text{DM}}) \diff^2 \vec{S} \diff  |\vec{v}_{\text{DM}}| \diff \Omega_v 
\end{align}
We will first evaluate the integral $\diff^2 \vec{S}$. For this integral we can keep the direction
of the DM velocity fixed and the scalar product is just the angle between
the two directions, $\theta_S$
\begin{align}
	R_{\text{total}} 
	&= \frac{  n_{\text{DM}} R_{\text{max}}^2  }{2} \int  |\vec{v}_{\text{DM}}|^3 | \cos \theta_S |  f_v(\vec{v}_{\text{DM}}) \diff \phi_S \diff \cos \theta_S \diff  |\vec{v}_{\text{DM}}| \diff \Omega_v  \nonumber\\
	&=  2 \pi \,   n_{\text{DM}} R_{\text{max}}^2  \int  |\vec{v}_{\text{DM}}|^3   f_v(\vec{v}_{\text{DM}})  \diff  |\vec{v}_{\text{DM}}| \diff \Omega_v \;.
\end{align}
For September 1, 2022, we find numerically
$R_{\text{total}}  = (13000 \text{ km/s} ) \times \,  n_\text{DM} \, R_\text{max}^2$.
For this paper we choose
\begin{equation}
n_{\text{DM}} = \frac{5.3\times 10^{-13}}{\text{km}^3}  \frac{1\text{ kg}}{m_\text{DM}} \;,
\end{equation}
corresponding to a DM energy density of $0.3$~GeV/cm$^3$.
The annual and daily modulation induced by the motion of the Earth relative to the galactic 
frame would modulate this number by a few percent and we will neglect the modulation for 
our final results.

\begin{figure}
	\centering
	\includegraphics[width=0.48\textwidth]{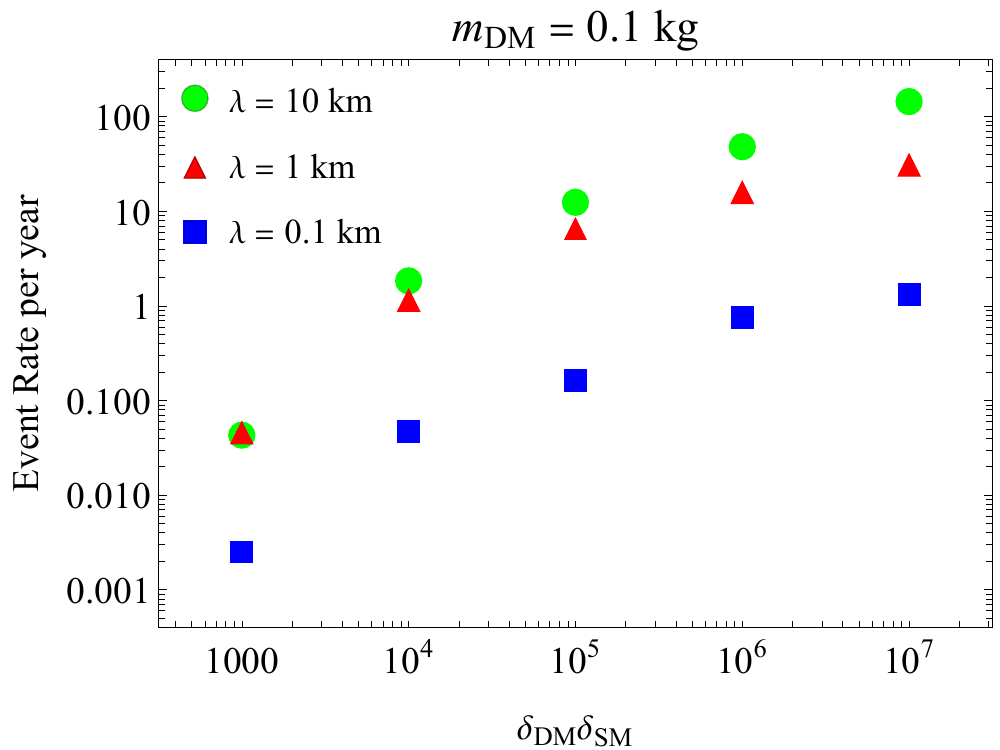} \hfill
	\includegraphics[width=0.48\textwidth]{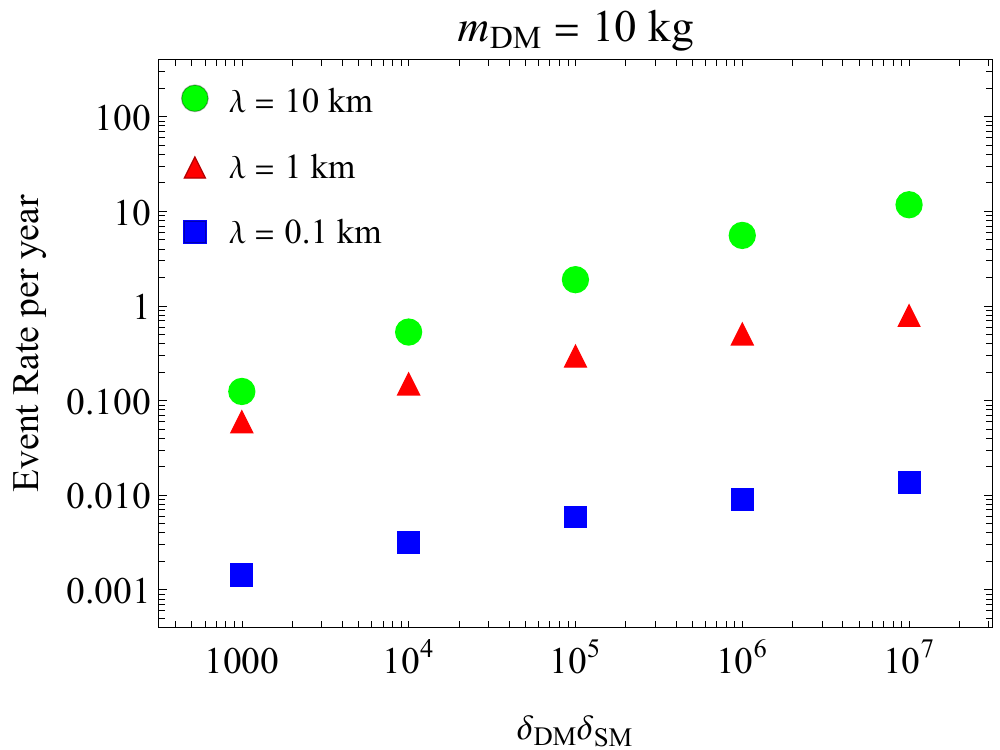}
	\includegraphics[width=0.48\textwidth]{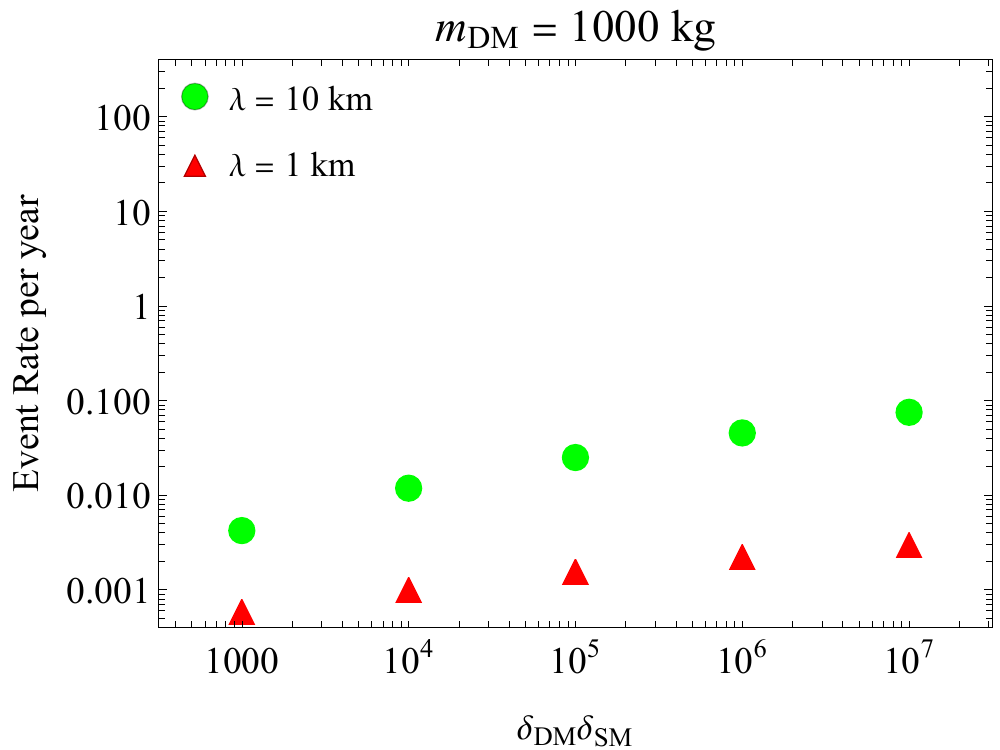}			
	\caption{
	   The expected rate for DM events with $\SNR > 1$ for various values of $m_\text{DM}$,
	   $\lambda$ and $\delta_{\text{SM}} \delta_\DM$ within our setup. For $m_\text{DM} = 1000$ kg and $\lambda = 0.1$ km, the rate is lower than $10^{-4}$ events per year and hence not shown in the figure.}
	\label{fig:rate}
\end{figure}

In Fig.~\ref{fig:rate}, we finally show the expected rate
for various values of $m_\text{DM}$, $\lambda$ and $\delta_{\text{SM}} \delta_\DM$.
The number of points and iterations is chosen such that the Monte Carlo integration error is 
smaller than the dot size in the figure.

From the plots we can see that lighter DM results in a bigger rate which is driven mostly by 
the larger number density of DM. For example, in the case of
$m_\text{DM} = 0.1$ kg, the rate can be larger than once per year
for $\lambda \gtrsim 1$ km and $\delta_{\text{SM}} \delta_\DM \gtrsim 10^4$.
If the DM is heavy enough, e.g., $m_\text{DM} = 1000$ kg, the number density
is too low to generate enough of a rate. This can be seen from Fig.~\ref{fig:boundary}
where we plot the $\rho_\text{KAGRA}^2 = 1$ boundary for various DM masses.
In the plot, the ratio between the area for $m_\text{DM} = 1000$ kg and
$m_\text{DM} = 10$ kg is around 2.57. However the number density of the heavier
dark matter drops by a factor of 100. Therefore the larger volume of the $\rho_\text{KAGRA}^2 = 1$
region cannot compensate for the decrease of the number density in the rate calculation.

For the other parameters ($\delta_{\text{SM}} \delta_\DM$ and $\lambda$), the boundary 
gets larger for larger values without any decrease of the number density. Hence the rate 
for those cases are higher for a larger value of parameters. For many parts of parameter 
space, we found that the rate is one event per year or higher. In those cases, KAGRA could 
be able to see or constrain dark matter given a few years of operation.

In Fig.~\ref{fig:rate}, we see that for a lower value of $\delta_{\text{SM}} \delta_\DM$
the distance between the points for different $\lambda$ are smaller compared to larger values
of $\delta_{\text{SM}} \delta_\DM$. For smaller values to get $\rho > 1$ the DM has to get very
close to the suspension systems and the exponential factor from the Yukawa potential,
cf.~Eq.~\eqref{eq:DMpotential}, is practically one. For larger values of 
$\delta_{\text{SM}} \delta_\DM$ on the other hand it is possible to get $\rho > 1$ for distances 
where the exponential factor becomes relevant. That affects how fast the fraction of significant 
DM events grows.

We also note here that KAGRA could not observe any events for DM with purely Newtonian
interactions, i.e., for $\delta_{\text{SM}} \delta_\DM = 0$ in the considered DM mass range.

\section{SUMMARY AND CONCLUSIONS}
\label{sec:Summary}

In this work we discuss how a GW detector can be utilized as an alternative technology to 
look for a certain kind of DM. Here, we assume DM has a gravitational and Yukawa long-range 
interaction with ordinary matter exert- ing an external force on the mirror suspension systems in 
the laser interferometer of a GW detector. The Yukawa force is assumed to be proportional to 
the DM and the target mass like the gravitational force. It is then possible to get relatively high 
strain amplitudes for heavy DM which can potentially overcome the background noise in the 
detector. Another important advantage of that assumption is that the long-range interaction leads 
to a large cross section which can overcome the small local DM number density and we can still 
expect reasonable event rates for a large portion of the possible parameter space.

We first model in detail the DM strain amplitude for KAGRA considering all four sets of suspension 
systems, one at each end of each interferometer arm. We compare it with the outcome from a 
single-mirror setup which is very similar to the case considered in our previous 
work~\cite{Lee:2020dcd}.  We see from our results that it is crucial to model all four suspension 
systems as the strains of the one- and four- mirror setup are considerably different depending on the 
initial conditions. To calculate the SNR from the DM signal strain one should naively integrate over 
a large frequency range which is numerically challenging. But we could show with a simple estimate 
and numerical checks that low frequencies especially below about 100 Hz in most cases dominate 
the SNR, in particular for SNR values around one.

It is obvious from its geometry that KAGRA cannot measure forces in all directions equally well. 
For that reason we had a closer look at its directional sensitivity for DM and studied how the SNR 
is affected by different DM velocities and initial positions. The SNR is obviously large when the DM 
would pass by close to one of the suspension systems. But we could also see that the effect of the 
force on the two arms can add up or cancel in the observable phase difference. Hence, KAGRA 
can even act as a directional DM detector although there are some degeneracies between 
different directions.

We then calculate the expected observable event rate which we define here as the rate of events 
with an SNR greater than one. We assume DM follows the standard halo model velocity 
distribution and then find that it is in principle possible for KAGRA to detect a few events per year 
for 10 kg DM, a Yukawa interaction with a range of 10 km, and an effective coupling constant 
around $10^{6}$. For lighter DM one could even expect more events since the number density 
increases faster than the cross section decreases. Hence, KAGRA and other GW detectors can 
serve as DM detectors for kg-scale DM with a long-range interaction, a possible scenario which is 
so far rather poorly constrained.

\section*{ACKNOWLEDGMENTS}

We would like to thank Alvina Yee Lian On for collaboration during early stages of this work and 
also for her comments on the manuscript. C. H. L. and M. S. are supported by the Ministry of 
Science and Technology (MOST) of Taiwan under Grants No. MOST 110-2112-M- 007-01 and 
No. MOST 111-2112-M-007-036. The work of R. P. is supported by the Parahyangan Catholic 
University under Grant No. III/LPPM/2022-02/72-P.

\end{document}